\documentclass{article}

\usepackage{adjustbox}
\usepackage{algorithmic}
\usepackage{amsmath,amssymb,amsfonts}
\usepackage{array}
\usepackage{authblk}
\usepackage{bibunits}
\usepackage{booktabs}
\usepackage{caption}
\usepackage{csvsimple}
\usepackage{etoolbox}
\usepackage{graphicx}
\usepackage{hyperref,cleveref}
\usepackage{import}
\usepackage{natbib}[sort]
\usepackage{rotating}
\usepackage[moderate]{savetrees}
\usepackage{siunitx}
\usepackage{subcaption}
\usepackage{textcomp}
\usepackage{xcolor}
\usepackage{lib/alifeconf}
\usepackage{orcidlink}
\usepackage{pdflscape,longtable}
\usepackage{listings}
\usepackage{placeins}
\usepackage{pythonhighlight}

\makeatletter
\let\pragma@iinput=\@iinput
\def\@iinput#1{\xdef\@pragmafile{#1}\pragma@iinput{#1} }
\def\@pragmafile{default}
\def\pragmaonce{%
   \csname pragma@\@pragmafile\endcsname
   \global\expandafter\let \csname pragma@\@pragmafile\endcsname =  
}
\makeatother

\ifdefined\mydraft
\mydraft
\fi

\defaultbibliography{bibl}
\defaultbibliographystyle{apalike-ejor}

\begin{document}

\title{ Methods to Estimate Cryptic Sequence Complexity }

\author[1,2,3,*]{Matthew Andres Moreno\orcidlink{0000-0003-4726-4479}}

\affil[1]{Department of Ecology and Evolutionary Biology, University of Michigan, Ann Arbor, United States}
\affil[2]{Center for the Study of Complex Systems, University of Michigan, Ann Arbor, United States}
\affil[3]{Michigan Institute for Data Science, University of Michigan, Ann Arbor, United States}
\affil[*]{corresponding author: \textit{morenoma@umich.edu}}

\maketitle

\begin{bibunit}

\begin{abstract}
\vspace{-1ex}
Complexity is a signature quality of interest in artificial life systems.
Alongside other dimensions of assessment, it is common to quantify genome sites that contribute to fitness as a complexity measure.
However, limitations to the sensitivity of fitness assays in models with implicit replication criteria involving rich biotic interactions introduce the possibility of difficult-to-detect ``cryptic'' adaptive sites, which contribute small fitness effects below the threshold of individual detectability or involve epistatic redundancies.
Here, we propose three knockout-based assay procedures designed to quantify cryptic adaptive sites within digital genomes.
We report initial tests of these methods on a simple genome model with explicitly configured site fitness effects.
In these limited tests, estimation results reflect ground truth cryptic sequence complexities well.
Presented work provides initial steps toward development of new methods and software tools that improve the resolution, rigor, and tractability of complexity analyses across alife systems, particularly those requiring expensive \textit{in situ} assessments of organism fitness.

% Our ultimate aim for this work is to improve the
% through development of rigorous, tractable
% We look forward to continuing work validating and making robust proposed methods.
% Ultimately, we hope to release methodology and accompanying tools that will help the community be able to more informatively study key questions about the evolution of complexity within sophisticated, i.

% Many artificial life systems have advantages due to their inherent interprability, where you can directly assess the meaning --- and the fitness relevancy --- of genetic sites.
% However, some systems for work on important topics like open-ended evolution, major transitions in evolution, and the interplay of biotic selection in ecologies are inherently more opaque and implicit.
% At the extremum of these systems, it becomes necessary to analyze genome content by studying in situ fitness effects of knockout variants.
% These tests, particularly when they must be performed as variant-versus-variant competitions, have inherent sensitivity limitations which mean that many sites are cryptic sequence complexity --- they contribute to fitness may not be detectable when individually knocked out.
% This may be due to small additive effects or to epistatic interactions (e.g., redundancy).
\end{abstract}
\vspace{-1ex}

\section{Introduction} \label{sec:introduction}

Understanding under what conditions and through what mechanisms complexity evolves is a significant open question in artificial life and evolutionary biology \citep{taylor2016open,pigliucci2009extended}.
Indeed, numerous dimensions of complexity in biological systems have been considered, including those of developmental processes, phenotypic traits, and ecological interactions \citep{szathmary2001can,mcshea2000functional}.
% Any reasonable treatment of the subject should consider multiple dimensions and not try to turn it into one dimension.
Among dimensions of biological complexity, the information content of genetic sequences is often useful due to its relative tractability, generalizability across systems, and foundational role to other aspects of biological complexity \citep{adami2002complexity}.
Although information theoretic formulations have been established to describe biological sequence complexity \citep{weiss2000information}, counts of adaptive sites (i.e., genome sites that benefit fitness) are a convenient, commonly-used proxy measure for genome sequence complexity \citep{dolson2019modes,moreno2021case}.

However, challenges can arise in identifying adaptive sites within a genome.
Beyond very small genome sizes, complete identification of adaptive sites is hindered by combinatoric effects that make all-combinations analyses necessary to fully detangle epistatic effects effectively intractable \citep{nitash2021information,adami2000evolution}.
This is particularly the case for systems with implicit fitness conditions with extensive biotic selection effects \citep{moreno2019toward,channon2000towards}.
In such circumstances, it can become necessary to use head-to-head competition trials between wildtype strains and knockout variants \textit{in situ} to detect fitness effects \citep{moreno2022exploring}.
Such competition-based fitness assays typically have sensitivity limitations, which limit detection of sites with small fitness effects.
Here, we term genome sites with adaptive effects that are not directly detectable through single-site knockouts as ``cryptic'' sequence complexity.

To enable more complete sequence complexity analyses of digital organisms inclusive of ``cryptic'' adaptive sites, we propose three assays to conduct statistical estimates of
\begin{enumerate}
\item \textbf{Additive-effect cryptic sites:} make small contributions to fitness that, individually, fall below the threshold of detectability of fitness assays;
\item \textbf{Epistatic-effect cryptic sites:} exhibit fitness effects only in the context of other knockouts (i.e., redundancy); and
\item \textbf{Any-effect cryptic sites:} make any contribution to fitness, inclusive of the above.
\end{enumerate}

The following sections describe proposed assays and report initial experiments with a simple model system designed to validate their estimations of cryptic sequence complexity.
Software, including full, documented implementations of underlying statistical estimators, is at \url{https://github.com/mmore500/cryptic-sequence-concept} \citep{moreno2024cryptic}.
% This project benefited from availability of many pieces of open-source software \citep{TODO}.

% \input{text/body/methods.tex}

\section{Additive-effect Cryptic Sites}

This assay is designed to detect sites with small-effect fitness contributions.
Although individual knockout effects of these sites do not reach the threshold for detectability, knocking out several may.
This assay assumes a preliminary set of single-site knockouts that exclude sites with individually detectable fitness effects from further consideration.

The assay proceeds by sampling sets of remaining sites are sampled and knocking them out together.
In the presence of small-effect sites, a classic dose-response curve will occur as knockout set size is increased.
That is, detectable fitness effects will be observed more frequently for larger knockout sets.
The shape of this dose-response curve will depend on both the abundance of small-effect sites and their mean effect size.
Fitting a negative binomial distribution to the curve allows these factors to be estimated.
This distribution models, for coin flips with success probability $p$, the number of successive trials required to achieve $n$ successes.
If we consider sampled sites being small-effect versus true-neutral as a coin flip event, then the proportion of small-effect sites in the genome will correspond to $p$ and the number of small-effect mutations required to reach detectability, $n$, will be inverse to the mean per-site effect size.
Under this framing, the dose-response curve corresponds to the cumulative distribution function of an underlying negative binomial distribution.

For an initial experiment exploring this approach, we generated a simple genome with 1,000 sites.
Fifty sites were designated to be small-effect sites, with effect sizes uniformly distributed between 0 and 0.7 (relative to a detectability threshold of 1.0).
To decide dose levels for the main assay (i.e., numbers of sites in knockout samples), we first sampled 250 doses spread evenly from 1 to 1,000 sites and tested for detectable fitness effects from one sample at each dose level.
We used the interval between the lowest dose with a detected fitness effect (25 sites) and the highest dose without a detected fitness effect (140 sites) as the dosing range for the main assay, choosing five dosing levels spaced evenly across this range.
We performed 1,000 knockouts of site sets sampled at each dosage level and then fit a negative binomial distribution to the observed outcome frequencies.
Resulting estimates of the additive site count and mean effect size were 30.0 and 0.5, approximating the true values of 50 sites and 0.35 effect size.
% https://github.com/mmore500/cryptic-sequence-concept/blob/004082f0cfe19dff8769aecbb3cc311be1db2a87/explore_additive.ipynb

\section{Epistatic-effect Cryptic Sites}

\begin{figure}
  \centering
  \footnotesize
  \includegraphics[width=\linewidth]{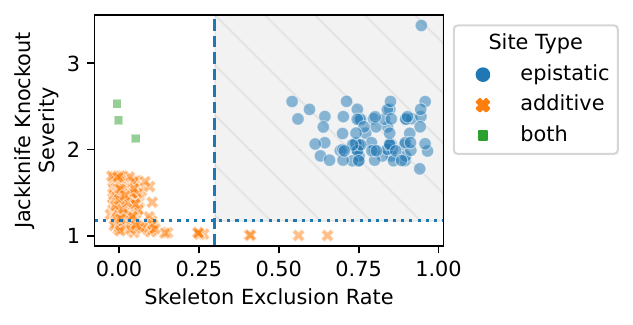}
  \vspace{-0.25in}
  \caption{%
    \textbf{Distinguishing between small-effect and epistatic genome sites.}
    \footnotesize
    Epistatic sites exhibit both 1) severe fitness effects when knocked out individually from ``skeletonized'' minimal viable genomes (i.e., ``jackknifed'') and 2) are often absent from sampled ``skeleton'' genomes (i.e., high exclusion rates).
    Minor jitter added to points for clarity.
  }
  \label{fig:epistatic}
  \vspace{-0.25in}
\end{figure}

This assay is designed to detect sites that only express detectable fitness knockout effects in the presence of specific other knockouts due to redundant masking.
For this assay we generated minimal fitness-equivalent genome ``skeletons'' by knocking out sites until no more could be removed without detectable fitness loss.
Because only one among a set of redundant sites will appear in any skeleton, the frequency with which redundant sites are \textit{excluded} from skeletons should be high.
However, some very small effect additive sites will also be able to be eliminated from skeletons.
To disambiguate these scenarios, we perform an additional step: ``jackknife'' one-by-one knockouts of each site within sampled skeletons.
Jackknifed sites with detectable, but very small magnitude fitness effects can then be identified as likely additive, rather than epistatic.

For an initial experiment exploring this approach, we generated a test genome with 4,000 sites, 200 of which were designated to be small-effect sites.
Effect sizes for these sites were uniformly distributed between 0 and 0.7, relative to the detectability threshold of 1.0.
We additionally introduced 20 sets of 5 redundant sites.
Fitness penalties uniformly distributed between 0.7 and 1.6 units were incurred when all sites within a set were knocked out.

To perform the assay, we first sampled 20 genome skeletons.
We then compared the frequency with which each site appeared in any skeleton to the severity of fitness effects when knocked out individually.
As Figure \ref{fig:epistatic} shows, epistatic sites are distinct in having both high skeleton exclusion rates and high jackknife knockout severity.
Applied to the sample genome, this procedure successfully identified 80 of the true 94 epistatic sites.

\section{Any-effect Cryptic Sites}

Any genome site with some fitness benefit should, in principle, potentially appear within a genome skeleton.
Put another way, skeletons sample randomly (though not uniformly) from genome sites that provide some fitness benefit.
The composition of genome skeletons can thus be analogized to the content of traps used by wildlife biologists to estimate the size of animal populations.
% In these scenarios, individuals tend to have lower recapture counts when diluted within a large population.
Robust statistical procedures exist to account for biasing factors, including variation in capture probabilities among individuals (``trap shyness'') \citep{amstrup2010handbook}.
We use the Burnham-Overton procedure to estimate the total number of any-effect sites from the distribution of capture counts among skeleton genome sites \citep{burnham1979robust}.

For an initial experiment exploring this approach, we generated a sample genome with 10,000 sites, 395 of which were designated to be small-effect sites, 155 were designated epistatic, and 5 were both small-effect and epistatic.
Effect sizes for the small-effect sites were uniformly distributed between 0 and 0.7, relative to a detectability threshold of 1.0.
Epistatic sites were organized into 40 4-site sets, with a fitness penalty between 0.7 and 1.6 units incurred when all sites within a set were knocked out.
We used progressive knockouts to sample 5 skeletons, with 504 distinct sites appearing in at least one skeleton.
The Burnham-Overton estimation procedure estimated 558 any-effect sites, close to the true count of 555 any-effect sites.
The 95\% confidence interval for this estimation was between 533 and 583 sites.

\section{Conclusion} \label{sec:conclusion}

Much work remains in developing proposed assays for cryptic sequence complexity.
It will be particularly critical to develop strategies to manage stochastic aspects of implicit fitness assays, a topic not treated here.
Another priority is formulation of assays to take advantage of parallel processing power, especially with respect to sequential operations like pruning out genome sites to produce a minimal viable skeleton.
The feedback-dependent workflows necessary to conduct these assays will likely necessitate development of software tools that orchestrate knockout trials and collate their results.
Future work should also assess statistical questions of how best to select maximally informative knockout targets and doses, as well as how bootstrapping or other procedures should be incorporated to provide confidence intervals for estimates of cryptic complexity.
% We will need methodologies to use parallel trials to rapidly strip down skeletons.
% And how to sequence knockouts performed for maximum statistical power.
Finally, building off initial, cursory demonstrations here, methods will need to be subjected to thorough and rigorous testing with a variety of full-fledged artificial life systems.
We look forward to these further steps on the path to improving the methodological capabilities of artificial life research for rigorous study of complexity.

\section*{Acknowledgment}

This work is supported by the Eric and Wendy Schmidt AI in Science Postdoctoral Fellowship, a Schmidt Futures program.

\putbib

\end{bibunit}

\clearpage
\newpage

% \begin{bibunit}

% \input{text/supplement.tex}

% \input{text/references.tex}

% \end{bibunit}

\end{document}